\documentclass[times,final,authoryear]{elsarticle}
\usepackage[margin=2.5cm]{geometry}
\usepackage[utf8]{inputenc}
\usepackage[english]{babel}
\usepackage{comment}
\usepackage{natbib}
\usepackage[utf8]{inputenc}
\usepackage[T1]{fontenc}
\usepackage[utf8]{inputenc}
\usepackage{babel}
\usepackage[font=small,labelfont=bf]{caption}
\usepackage{multirow}

\usepackage{mathtools}

\usepackage{amssymb}
\usepackage{amsmath}
\newcommand{\R}{\mathbb{R}}

\usepackage{fixltx2e}

\usepackage{floatrow}
\newfloatcommand{capbtabbox}{table}[][\FBwidth]

\usepackage{blindtext}
\usepackage{amsmath,amssymb,mathtools,amsfonts,amsthm,xcolor}

\def\hat{\widehat}

\def\Jb{{\bf J}}

\def\Hb{{\bf H}}
\def\Xb{{\bf X}}
\def\Wb{{\bf W}}
\def\xb{{\bf x}}
\def\Yb{{\bf Y}}

\def\bb{{\bf b}}

\def\rb{{\bf r}}

\def\bb{{\bf b}}

\def\thetab{\boldsymbol \theta}
\def\betab{\boldsymbol \beta}
\def\deltab{\boldsymbol \delta}
\def\mub{\boldsymbol \mu}

\usepackage{listings}
\usepackage{float}
\usepackage{hyperref}

\usepackage{subfig}

\usepackage[T1]{fontenc}
\usepackage[font=small,labelfont=bf,tableposition=top]{caption}
\usepackage[T1]{fontenc}
\usepackage[utf8]{inputenc}
\usepackage{babel}
\usepackage[font=small,labelfont=bf]{caption}
\DeclareCaptionLabelFormat{andtable}{#1~#2  \&  \tablename~\thetable}
\journal{International Encyclopedia of Education 4th Edition}

\begin{document}

\begin{frontmatter}


\title{Nonlinear Regression Analysis}



\author{Hsin-Hsiung Huang and Qing He}

\address{Department of Statistics and Data Science, University of Central Florida, United States}

\begin{abstract}
Nonlinear regression analysis is a popular and important
tool for scientists and engineers. In this article, we introduce theories and methods of nonlinear
regression and its statistical inferences using the frequentist and Bayesian statistical modeling and computation.
Least squares with the Gauss-Newton method is the most widely used approach to parameters estimation. Under the assumption of normally distributed errors, maximum likelihood estimation is equivalent to least squares estimation. The Wald confidence regions for parameters in a nonlinear regression model are affected by the curvatures in the mean function. Furthermore, we introduce the Newton-Raphson method and the generalized least squares method to deal with variance heterogeneity.  Examples of simulation data analysis are provided to illustrate important properties of confidence regions and the statistical inferences using the  nonlinear least  squares estimation  and Bayesian inference.
\end{abstract}


\end{frontmatter}

\section{Introduction}
\label{S:1}

\subsection{An overview of the nonlinear regression models}
Nonlinear regression models have been widely used in various fields including statistics, chemistry, physics, psychology, health science, and biology. 
Some of them have a linear relationship in the parameters (i.e., linear in the $\thetab$). For example, a polynomial regression fits a curved relationship between the response variable and predictors using higher-ordered values of the predictors, but it is linear in terms of the parameters. Statistical inference can be derived from nonlinear regression models~\eqref{nonlinear_model}. 
The main advantages of nonlinear regression models include interpretability, parsimony, and prediction \citep{Bates_Watts1988}. In general, nonlinear models are capable of accommodating various mean functions, so nonlinear models are appropriate for applications with parsimonious parameters  and easily interpretable due to the fact that the parameters can be associated with meaningful factors. 

Assume that there are $n$ observations of responses $y_1, \ldots ,y_n$ and predictors $\xb_{ik}$ associated with $y_i$ on $k=1,\ldots,p$ independent variables. In nonlinear regression, the $y$'s and $\xb$'s satisfy the nonlinear
regression model
\begin{equation}
    y_i = f(\xb_i,\thetab) + \epsilon_i,\; i= 1,\ldots, n,
    \label{nonlinear_model}
\end{equation}
where $f$ is a function that is nonlinear in the $p$ elements of parameter $\thetab$ with the unknown true value $\thetab^*$, and the $\epsilon_i$'s are independent and identically distributed (i.i.d) random errors with mean zero
and constant variance $\sigma^2$.
The conditional
mean response $\mbox{E}(y_i|\xb_i)=f(\xb_i,\thetab)$, so we will refer
to $f$ as the mean function.
For example, the Beverton-Holt model \citep{beverton2012dynamics} which 
is similar to the Michaelis-Menten model \citep{sheiner1980evaluation} and used for modeling discrete-time population  gives the expected population $y=n_{t + 1}$ as a function of the previous population $\xb=n_t$ as
$$
n_{t+1}=f\left(\xb,(\alpha,\beta)\right)=\frac{
\alpha \xb}{1 + \xb/\beta},
$$ 
where $\thetab=(\alpha,\beta)$ and the parameter $\alpha$ is the slope at $0$ and $\beta$ is the concentration between $0$ and the upper limit $\alpha\beta$. It is a nonlinear regression since the two variables are related in a nonlinear (curved) relationship.

\subsection{Estimation methods}
In order to estimate the parameters in a nonlinear regression model, we minimize the sum of squared residuals, which measures how far the $y$ observations vary from the nonlinear function $f(\xb_i,\thetab)$ used to predict $y$.
Such estimators are called the least squares estimators $\hat{\thetab}$ of $\thetab$,
that is,
$$
S(\thetab) =\sum^n_{i=1}
(y_i - f(\xb_i,\thetab))^2$$
is used to find
\begin{equation*} \hat{\thetab}=\arg\min_{\thetab}S(\thetab). \end{equation*}

For example, to estimate $\alpha$ and $\beta$ in the Beverton-Holt model, we minimize $S(\alpha,\beta) = \sum(y_i - \frac{\alpha \xb_i}{1+\xb_i/\beta})^2$. We first derive the gradients $g_\alpha, g_\beta$ and Hessian matrix with elements $H_{i,j},\; i,j\in\{\alpha,\beta\}$ as follows:
\begin{align*}
g_\alpha &=\frac{ \partial{S(\thetab)}}{\partial{\alpha}}=-2\sum{\xb_i (y_i-\frac{\alpha \xb_i}{1+\xb_i/\beta})},\\
g_\beta &=\frac{ \partial{S(\thetab)}}{\partial{\beta}}=-2\sum{(y_i-\frac{\alpha \xb_i}{1+\xb_i/\beta}) \frac{\alpha \xb_i^2}{\beta^2(1+\xb_i/\beta)^2}},\\
H_{\alpha,\alpha}&=\frac{ \partial{g_\alpha}}{\partial{\alpha}}=2\sum{\frac{ \xb_i^2}{1+\xb_i/\beta}},\\
H_{\alpha,\beta}&=\frac{ \partial{g_\alpha}}{\partial{\beta}}=2\sum{\frac{\alpha \xb_i^3}{\beta^2(1+\xb_i/\beta)^2}},\\
H_{\beta,\alpha}&=-2\sum{\frac{ \xb_i^2 y_i}{\beta^2(1+\xb_i/\beta)^2}-\frac{\xb_i^3}{\beta^2(1+\xb_i/\beta)^3}},\text{ and }\\
H_{\beta,\beta}&=-2\sum{\frac{-2 \alpha\xb_i^2 y_i}{(\beta+\xb_i)^3}+\frac{2 \alpha^2 \xb_i^3}{\beta^3 (1+\xb_i/\beta)^3 } - \frac{3 \alpha^2 \xb_i^4}{\beta^4 (1+\xb_i/\beta)^4 }}.\\
\end{align*}
Then we can use the Gauss-Newton or Newton-Raphson methods to iteratively estimate $\alpha$ and $\beta$ introduced in Section~\ref{sec:GN} and \ref{sec:NR}, respectively. The Gauss-Newton method is essentially the Newton-Raphson method with the modification that the Gauss-Newton method uses the approximation $2J^TJ$, where $J$ is the Jacobian matrix, for the Hessian matrix. Note that when for a scalar nonlinear function its Jacobian matrix is same as the gradient.

\section{Gauss-Newton method for least squares estimation}\label{sec:GN}

When $f$ is twice differentiable with respect to $\thetab$, 
we have the gradient and Hessian which can be written as:
\begin{align*}
g_j &= \frac{\partial S(\thetab)}{\partial \theta_j}
= - 2 \sum^n_{i=1}(y_i-f_i) \frac{\partial f_i}{\partial \theta_j},\\
H_{jk} &= \frac{\partial^2 S(\thetab)}{\partial \theta_j\partial \theta_k}
= - 2 \sum^n_{i=1}\left((y_i-f_i) \frac{\partial^2 f_i}{\partial \theta_j \partial \theta_k}-\frac{\partial f_i}{\partial \theta_j }\frac{\partial f_i}{\partial \theta_k}\right)
\end{align*}
where $f_i = f(\xb_i, \thetab)$.

Further, we solve
for the least squares solution $\hat{\thetab}$ in the following
system of equations
\begin{align}
\left.\frac{\partial S(\thetab)}{\partial \theta_j}\right|_{\thetab=\hat{\thetab}}
= 0; \; j= 1,\ldots, p,
\label{normal_equ1}
\end{align}
which are called the normal equations with
$$
\left.\sum^n_{i=1}(y_i-f_i) \frac{\partial f(\thetab)}{\partial \theta_j}\right|_{\thetab=\hat{\thetab}}=0.
$$
Its matrix form is
$$
\Jb(\hat{\thetab})^T\rb= 0
$$
with $(i, j)$-th element  $J_{i,j} (\hat{\thetab})= \frac{\partial f_i}{\partial \theta_j}$ and $\rb =
y- f(\hat{\thetab})$ with the $i$-th element $r_i = y_i - f_i$.
The matrix $\Jb (\hat{\thetab})$ of size $n\times p$ is called the Jacobian matrix. The
$p \times p$ matrices $G_1,\ldots, G_n$ so that the $(j, k)$-th element of $G_i$ is $\frac{\partial^2 f_i}{\partial \theta_j \partial \theta_k}$, and then the gradient and Hessian are given by
\begin{align}
g = -2\Jb^T \rb,\; \Hb = 2\Jb^T \Jb - 2
\sum^n_{i=1}
r_iG_i.
\end{align}
In a linear regression model,
the Jacobian is simply the data design matrix $\Xb$ and does not depend on the parameter values $\thetab$.

The normal equations do not have an analytic
solution for $\hat{\thetab}$ in most cases, so that numerical iterative procedures are needed. The Gauss-Newton method estimates the parameters in nonlinear regression using the first-order Taylor's expansion 
$$f(\xb,\thetab)\approx f(\xb,\thetab^*)+\nabla f(\xb,\thetab)^T(\thetab-\thetab^*)$$
to approximate a nonlinear regression function \citep{Nada:1964,Rupp:Wand:1994} considering a small neighborhood of $\thetab^*.$ The linear approximation of $f(\xb,\thetab)$
in the neighborhood of $\thetab^*$ results in an approximate residual sum of squares
$$
S(\thetab)
\approx
\sum^n_{i=1}
\left(y_i - f(\xb_i, \thetab^*) 
-\sum^p_j J_{i,j}^T(\theta_j-\theta_j^*)
\right)^2,
$$
and 
\begin{equation}
S(\thetab)-S(\thetab^*)
\approx   (\hat \thetab - \thetab^*)^T\Jb^T\Jb(\hat \thetab - \thetab^*).
    \label{approx::res_diff}
\end{equation}
The corresponding normal equations at the
$t$-th iteration are given by
$$\Jb(\thetab^{(t)})^T \Jb(\thetab^{(t)})(\thetab-\thetab^{(t)}) = \Jb(\thetab^{(t)})^T(y- f(\thetab^{(t)})).$$
The updating increment
\begin{equation*}
   \deltab^{(t)} = \left(\Jb(\thetab^{(t)})^T \Jb(\thetab^{(t)})\right)^{-1}\Jb(\thetab^{(t)})^T(y- f(\thetab^{(t)})) 
\end{equation*}
in the $t$-th iteration, and then $\thetab^{(t+1)}=\thetab^{(t)}+\lambda_t\deltab^{(t)}$
where $\lambda_t$ is the step size. The update direction $\deltab^{(t)}$
is derived from the tangent plane approximation
to the solution locus.

The estimate of the asymptotic covariance of the
least squares estimate is given by
$$
\mbox{cov}(\hat{\thetab}) = \sigma^2(\Jb^T \Jb)^{-1}.$$ The statistical inference about $\thetab$ is based on the property that
$\hat{\thetab}$ follows $N(\thetab, \sigma^2(\Jb^T \Jb)^{-1})$ asymptotically. $\sigma^2$ is estimated by $s^2 = S(\hat{\thetab})/(n-p)$, where $p$ is the number of parameters and $\Jb$ is estimated by $\Jb(\hat{\thetab})$.

\subsection{Maximum likelihood estimation}

Considering a normal distributed error term
$$
y_i - f(\xb_i, \thetab)  \stackrel{i.i.d.}{\sim} N(0,\sigma^2),
$$
the log-likelihood function of $\thetab$ and $\sigma^2$ is
\begin{align*}
l(\thetab, \sigma^2) = -\frac{S(\thetab)}{2 \sigma^2}  - 
\frac{n}{2}\log (2\pi\sigma^2).
\end{align*}

Maximum likelihood estimation (MLE) is a method of estimating the parameters by maximizing 
$l(\thetab, \sigma^2)$ with respect to $\thetab$ and $\sigma^2$ for which we need the following derivatives
\begin{align}
\frac{\partial l}{\partial \sigma^2} &=
\frac{1}{2\sigma^4}
S(\thetab) - \frac{n}{2\sigma^2},\label{normal_equ2}\\
\frac{\partial l}{\partial \thetab} &=
-\frac{1}{2\sigma^2}\frac{\partial S(\thetab)}{\partial \thetab}.
\label{normal_equ3}
\end{align}
Setting equations~\eqref{normal_equ2}, \eqref{normal_equ3} equal to $0$, we have $\hat{\sigma}^2_{\footnotesize\mbox{MLE}} = \frac{S(\hat{\thetab})}{n}$
and the normal equation~\eqref{normal_equ1}. As a result, $
\hat{\thetab}_{\footnotesize\mbox{MLE}} = \hat{\thetab}.$
In summary, when the noise follows a normal distribution, the Gauss-Newton iterations to minimize the least squares $S(\thetab)$ are exactly the same as the Newton-Raphson iterations to maximize the log-likelihood function with respect to $\thetab$, and
hence the maximum likelihood estimator is equivalent to the least squares estimator \citep{Bates_Watts1988}.
\subsection{Asymptotic statistical inference}
Let us consider the nonlinear regression model~\eqref{nonlinear_model}
where the $\epsilon_i\stackrel{i.i.d}{\sim}  N(0,\sigma^2)$. We note that as $n\to\infty$, the least squares estimate $\hat{\thetab}$ is asymptotically
$N\left(\thetab^*,\sigma^2(\Jb^T\Jb)^{-1}\right)$, where $\thetab^*$ is the true value of $\thetab$, and the matrix
$\Jb= [ (\partial f(x_i ; \thetab ) /\partial \theta_j ]_{\thetab^*}$ plays the same role as $X$ in the linear regression \citep{saber1989nonlinear}. In particular, by analogy with the linear confidence region
\begin{equation}
  \{\thetab\;|\; (\thetab-\hat{\thetab})\hat{\Jb}^T\hat{\Jb}(\thetab-\hat{\thetab})\leq p s^2 F^{\alpha}_{p,n-p}\},  \label{asym:CR}
\end{equation}
is an approximate $100(1 - \alpha)\%$ confidence region for $\thetab$. Here $s^2 = S(\hat{\thetab})/(n - p)$,
$\hat{\Jb} = \Jb(\hat{\thetab})$, and $F^{\alpha}_{p,n-p}$ is the upper a critical value of the $F_{p,n-p}$ distribution. As
the linear approximation is valid asymptotically, \eqref{asym:CR} will have the correct
confidence level of $1 -\alpha$ asymptotically. As $\alpha$ varies, the regions \eqref{asym:CR} are
ellipsoid contours of the approximate multivariate
normal density function of $\hat{\thetab}$ (with $\Jb$ replaced by $\hat \Jb$). Since $S(\thetab)$ measures how the observations are close to the fitted equation for any $\thetab$, it would seem
appropriate to also base confidence regions for $\thetab$ on the contours of $S(\thetab)$. Such
a region could take the form
$$
\left\{\thetab\;|\; S(\thetab)\leq cS(\hat \thetab)\right\}
$$
for some $c$, $c > 1$. Regions of this type are often called the exact confidence regions,
as they are not based on any approximations. However, the confidence levels,
or coverage probabilities, of such regions are generally unknown, though
approximate levels can be obtained from asymptotic theory.

For large enough $n$, the consistent estimator $\hat \thetab$ will be sufficiently close to $\thetab^*$
for the approximation \eqref{approx::res_diff} to hold with $F^{\alpha}_{p,n-p}$ in \eqref{asym:CR}. Therefore,
substituting
\begin{align}
    S(\thetab^*) - S(\thetab)\approx (\hat \thetab  -\thetab^*)^T\hat \Jb^T \hat \Jb(\hat \thetab  -\thetab^*) 
    \label{approx::res_diff2}
\end{align}
from \eqref{approx::res_diff}, we obtain the confidence region 
\begin{align}
   \left\{\thetab\;|\; S(\thetab) \leq S(\hat \thetab)\left ( 1+\frac{p}{n-p}F^{\alpha}_{p,n-p}\right) \right\}
    \label{approx::CR}
    \end{align}
that has the required asymptotic confidence level of $100(1 -\alpha )\%$. The regions \eqref{asym:CR} and \eqref{approx::CR} are
asymptotically the same, and they are identical for
linear models. However, for finite $n$, these regions may be very different, and it indicates inadequacy of the linear approximation
\eqref{approx::res_diff}. 
Under the assumption of normal errors we see that the above clash between
the two confidence regions embodies a wider principle. We have a choice
between confidence regions based on the asymptotic normality of the maximum likelihood
estimator $\hat \thetab$, which is also the least squares estimator,
and those based on the contours of the likelihood function via the likelihood ratio
test.



\section{Parameterization and curvature}
There are some nonlinear models which are intrinsically linear because they can be made linear in the parameters by a simple transformation. For example, the reciprocal of the Michaelis-Menten regression mean function
\begin{equation} f(x,\bb)=\frac{b_{1}x}{b_{2}+x}
\label{MM-model}
\end{equation}
results in the model
\begin{align*} \frac{1}{y}&=\frac{1}{b_{1}}+\frac{b_{2}}{b_{1}}\frac{1}{x}+\epsilon\\ &=\beta_{1}+\beta_{2}\frac{1}{x}+\epsilon, \end{align*}
which is linear in the transformed parameters 
$b_1$
 and 
$b_2$. In such cases, transforming a model to its linear form often provides better inference procedures and confidence intervals. However, parameterization may affect the asymptotic variance and convergence of the estimation algorithm \citep{Bates_Watts1988, Huang2010}.
The re-parameterization is an important issue. Although the shape of the solution locus is fixed, the performance of the Gauss-Newton estimator varies with respect to different parameterizations. This depends on the nonlinearity
in the model. There are two kinds of nonlinearity--intrinsic nonlinearity and parameter-effects nonlinearity \citep{Hamilton1985,Bates_Watts1988}. The intrinsic nonlinearity is associated with the modeling and is invariant under parameterization. The parameter-effects nonlinearity, however, can be lessened through a proper parameterization. If either component of the nonlinearity is large, the least squares estimate may not converge. Furthermore, the estimated covariance for $\hat{\bb}$ given by 
$\hat{V}(\hat{\bb})=\hat{\sigma}^2
\left(\Jb (\hat{\bb})^T\Jb (\hat{\bb})\right)^{-1}
,$
would change greatly in each step of the iteration, and the statistical inference based on the asymptotic normality becomes unreliable. In other
words, the least squares estimator in nonlinear regression depends on the curvedness of the underlying model as well as the parameterization adopted.

The quality of the linear approximation can be summarized by intrinsic curvature
$\kappa^N_d = \|d^TC_Nd\|$
and parameter-effects curvature $\kappa^T_d = \|d^TC_Td\|$ \citep{Bates_Watts1988}, where $d$ has unit length, $C_N$ is the normal direction curvature matrix, and $C_T$ is the tangent direction curvature matrix. For details on curvatures, please read the Appendix in \cite{Huang2010}.

The two curvatures 
correspond to two assumptions of  linear approximation of a nonlinear mean function. First, the planar assumption ensures that the nonlinear mean function is approximated by its tangent plane at a given point. Second, the uniform coordinate assumption means that an equi-spaced, straight and parallel linear coordinate system is placed on the approximation
tangent plane. Intrinsic curvature is related to the planar assumption, and it depends on the dataset considered and the mean function but not on the parameterization used in the mean function. Parameter-effects curvature is related
to the uniform coordinate assumption, and it depends on all aspects of the
model, including the parameterization. Large values of these two curvature
measures indicate a poor linear approximation. The function rms.curv() in the R package MASS can be used to calculate the two measures for a given nls()
model fit \citep{Mass2002}. The intrinsic curvature is generally relatively negligible compared with the parameter-effects curvature \citep{Bates_Watts1988}, and in such cases the profile likelihood approach is useful.

The Wald, likelihood ratio (LR) and score tests are asymptotically equivalent. They all have the same sampling distribution which is a chi-square distribution with the same degrees of freedom. However, the Wald test statistics $(\hat{\bb}-\bb_0)^T \left[ \hat{V}(\hat{\bb})\right]^{-1} (\hat{\bb}-\bb_0)$, which results in an ellipsoidal confidence region, uses $\hat{\bb}$ and depends on curvature of the likelihood at $\hat{\bb}$. The score test depends on the slope and curvature at $\hat{\bb}_0$, while the LR test uses information from both $\hat{\bb}$ and $\hat{\bb}_0$. The differences among them vanish in large samples if the null hypothesis is true. If the null hypothesis is false, they may take very different values. 
The Wald procedures are influenced by both intrinsic curvature and parameter-effects curvature. A model with a lower parameter-effects curvature is
preferable.

\begin{figure}[ht]%
    \centering
    \subfloat[\centering n=50 ]{\includegraphics[width=5cm]{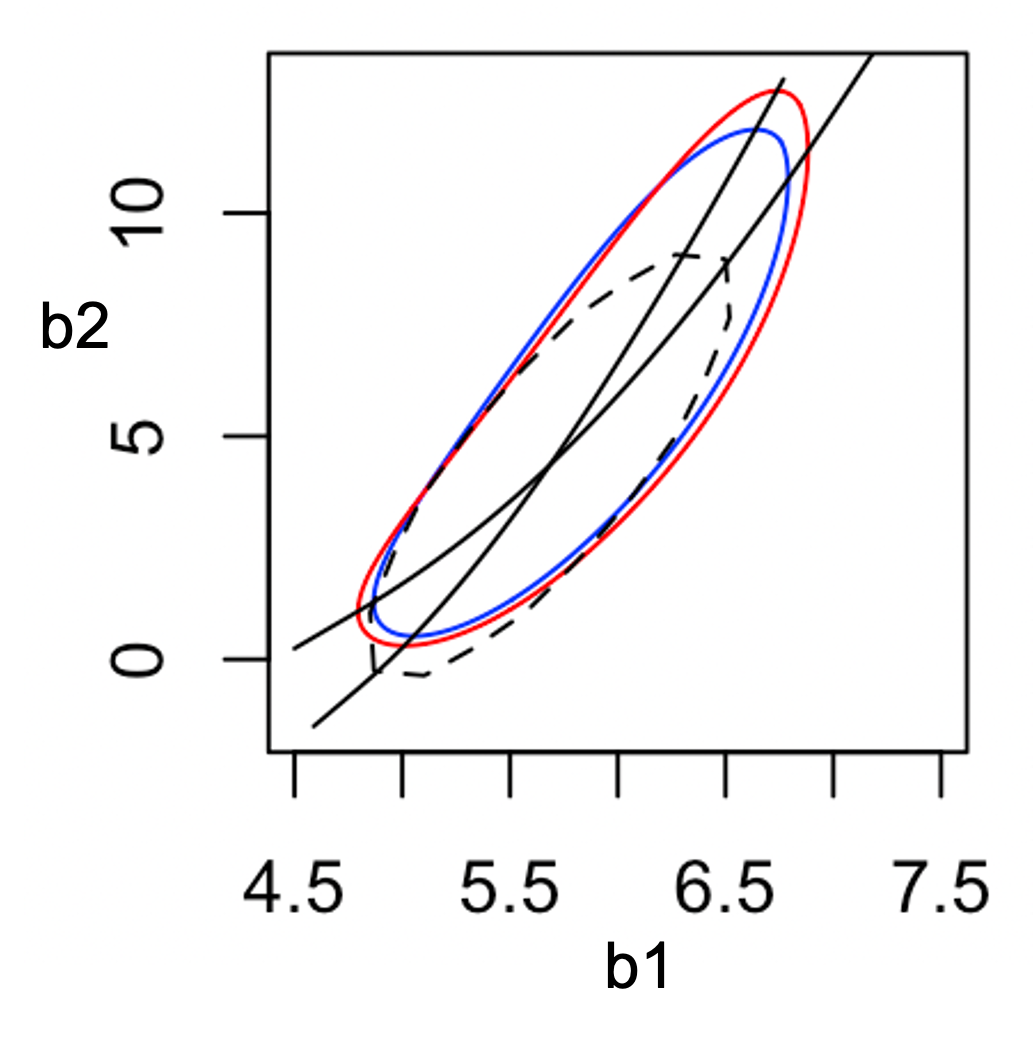}}
        \qquad
    \subfloat[\centering n=100
    ]{{\includegraphics[width=5.2cm]{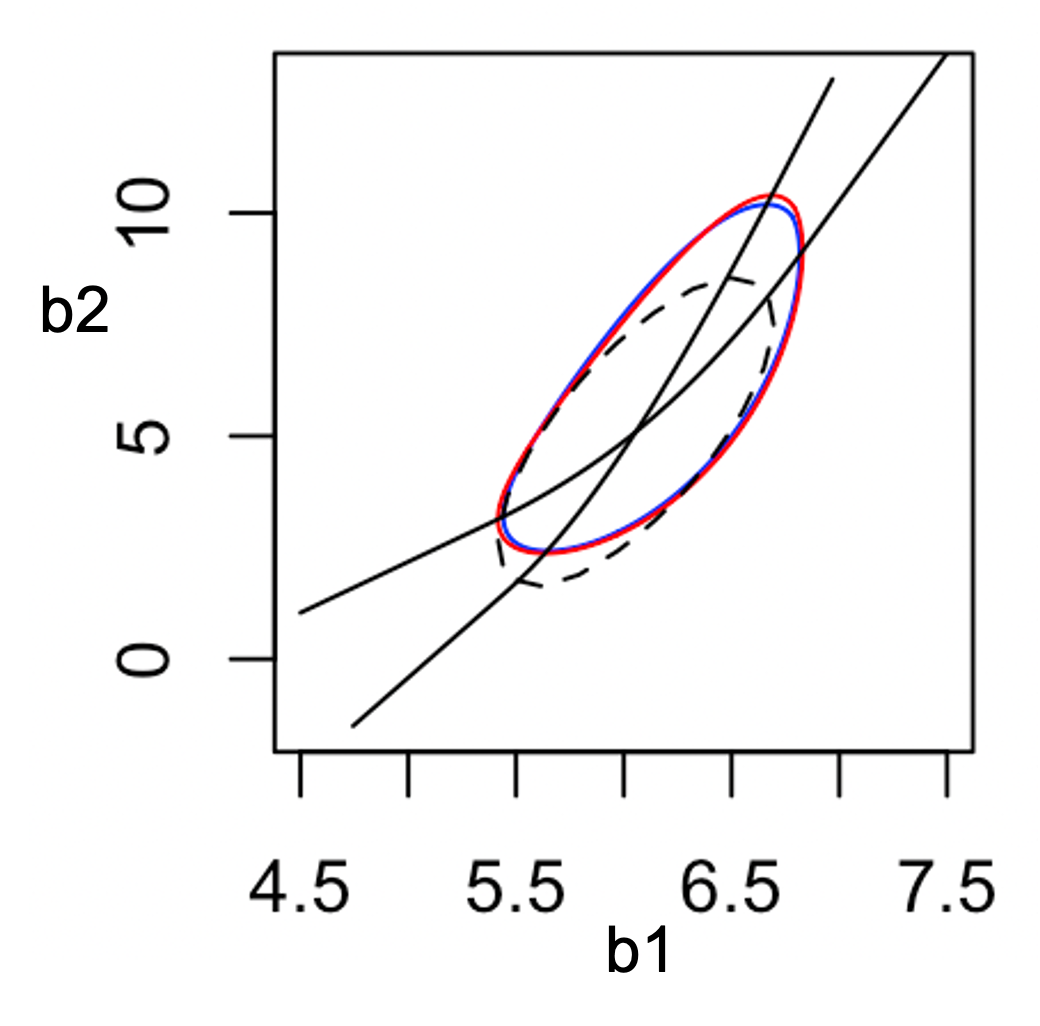} }}
    \caption{The profile plots of the simulation data analysis. The $95\%$ confidence region of $b_1$ and $b_2$ using the Wald statistic (dashed), the likelihood ratio (blue) and Skovgaard’s approximation (red) with simulated data from the Michaelis-Menten model. The left panel is with sample size $n = 50$, and the right panel is with sample size $n = 100$.}%
    \label{fig:calcium}%
\end{figure}


The curvature can affect the quality of statistical inferences based on the asymptotic normality. When the mapped parameter
curves onto the tangent plane are not uniform grid
lines, the resulting confidence region may not be reliable. We use the Michaelis-Menten
model~\eqref{MM-model} as an example by simulating data with  $\epsilon_i \stackrel{i.i.d.}{\sim}N(0,1^2)$, $ b_1 =
6, b_2 = 5$, and $\xb$ generated from the uniform distribution $(1, 100)$. We demonstrate how confidence regions of $b_1$ and $b_2$ are affected by curvatures with two different sample sizes $n_1 = 50$ and $n_2 = 100$ displayed in Figure~\ref{fig:calcium} using the contour method of the nlreg package \citep{nlreg}. The   dashed, blue and red lines represent the confidence regions of $(b_1,b_2)$ obtained with the Wald test, the likelihood ratio test, and Skovgaard's approximation, respectively \citep{skovgaard1996}. The confidence regions of the latter two methods are closer to the exact one than the Wald \citep{Bates_Watts1988,skovgaard1996}. When the sample size is small, the shapes and coverage between
the Wald confidence region and the two others are very different. As the sample size grows, they become less different.

\section{Newton-Raphson Method}\label{sec:NR}

The Newton-Raphson method (also called Newton's method) which has been widely used for estimating the parameters of nonlinear regression models due to its advantage of fast convergence using the second order Taylor's expansion to approximate the nonlinear function $l(\thetab)$ as a quadratic polynomial as follows:
\begin{align}
l(\mathbf{\thetab}) \approx l(\mathbf{\thetab}^{(t)}) + (\mathbf{\thetab} - \mathbf{\thetab}^{(t)})^T\nabla l(\mathbf{\thetab}^{(t)})  + \frac 12 (\mathbf{\thetab} - \mathbf{\thetab}^{(t)})^T \nabla^2 l(\mathbf{\thetab}^{(t)})  (\mathbf{\thetab} - \mathbf{\thetab}^{(t)}),
\label{NR:1}
\end{align}
where $\thetab^{(t)}$ is in a neighborhood of the true $\thetab$, $\nabla$ is the first derivative and $\nabla^2$ is the second derivative with respect to $\thetab$.
The Newton-Raphson method can be computationally challenging and heavily dependent on good starting values \citep{gill1986user}.
To maximize the quadratic approximation, we set the first derivative of the approximate quadratic function \eqref{NR:1} equal to $0$ at a new point $\thetab^{(t+1)}$
\begin{align*}
\nabla l(\thetab^{(t)}) + [\nabla^2 l(\thetab^{(t)})] (\thetab^{(t+1)} - \thetab^{(t)}) = \mathbf{0}_p,
\end{align*}
\begin{align*}
\thetab^{(t+1)} &= \thetab^{(t)} - [\nabla^2 l(\thetab^{(t)})]^{-1} \nabla l(\thetab^{(t)}) \\
&= \thetab^{(t)} + [-\nabla^2 f(\thetab^{(t)})]^{-1} \nabla f(\thetab^{(t)})
\end{align*}
which approximates $\thetab$ in each iteration $t$.
The value $\thetab^{(t+1)}$ should be a better guess than the initial point $\thetab^{(t)}$ is. 

Here, we focus on deriving the Newton-Raphson method for the exponential family \citep{Neld:Wedd:1972} including the Gaussian (normal) distributions for continuous variables, binomial distributions for binary response variables, Poisson distributions for count variables, and the gamma and inverse-Gaussian families of distributions for modeling positive continuous data, where the conditional variance of $Y_i$ increases with its expectation.
The log-likelihood for $n$ independent observations from an exponential family is given by
\begin{align}
L_n(\thetab,\phi;Y) = \sum_{i=1}^{n}{\left\{ \frac{y_i\theta_i - b(\theta_i)}{a_i(\phi)} + c(y_i,\phi) \right\}}.
\label{log_like}
\end{align}
Assume that the mean $\mu_i$ depends linearly on the covariates through a link function $g$ as follows:
$$
g(\mu_i)=\eta_i=X_i^T \betab .
$$
Because the link function is invertible, we can also write
$$
\mu_i=g^{-1}(\eta_i)=g^{-1}(X_i^T \betab) .
$$
Note that $\mu_i=b^{'}(\theta_i), \theta_i=(b^{'})^{-1}\circ g^{-1}(X_i^T \beta):=h(X_i^{T}\betab)$ where
\begin{align*}
   b^{'}(\theta_i)=\frac{\partial b(\theta_i)}{\partial \theta_i},\; g^{'}(\mu_i)=\frac{\partial g(\mu_i)}{\partial \mu_i},\;
   \text{and } h^{'}(\betab)=\frac{\partial h(\betab)}{\partial \betab}. 
\end{align*}

According to the chain rule, we have
\begin{align*}
\frac{\partial L_n}{\partial \beta_j}&=\sum_{i=1}^{n}\frac{\partial L_i}{\partial \theta_i}\frac{\partial \theta_i}{\partial \beta_j}\\
&=\sum_{i}^{n}\frac{Y_i-\mu_i}{\phi}h^{'}(X_i^T \betab) X_{i,j}\\
&:=\sum_{i}^{n}(Y_i-\mu_i)W_i X_{i,j},
\end{align*}
where $W_i = h^{'}(X_i^T \betab)/g^{'}(\mu_i)\phi$.\\
Define $\Wb=\text{diag}(W_1,\ldots,W_n)$, then the gradient is\\
\begin{align*}
\nabla L_n(\betab) = \sum_{i=1}^n \frac{(Y_i-\mu_i) \mu_i'(\eta_i)}{\sigma_i^2} X_i=\Xb^T \Wb (\Yb-\mub).
\end{align*}
For the Hessian, we have
$$
\frac{\partial^2 L_n}{\partial \beta_j\beta_k} = \sum_{i}\frac{Y_i-\mu_i}{\phi} h^{''}(X_i^T \betab) X_{i,j} X_{i,k}
-\frac{1}{\phi}\sum_i\left(\frac{\partial \mu_i}{\partial \beta_k}\right) h^{'}(X_i^T \betab) X_{i,j}.
$$
Note that 
$$
\frac{\partial \mu_i}{\partial \beta_k}=\frac{\partial b^{'}(\theta_i)}{\partial \beta_k}=\frac{\partial b^{'}(h(X_i^T \betab))}{\partial \beta_k}=b^{''}(\theta_i)h^{'}(X_i^T \betab)X_{i,k}.
$$
So, 
$$
-E[\Hb_{L_n(\betab)}]=\Xb^T \Wb \Xb,
$$
where $\Wb=\text{diag}\left(\frac{h^{'}(X^T \betab)}{g^{'}(\mu_i)}\right)$.
It leads to the Hessian matrix
 $\Hb(\thetab)=-\Xb^T\Wb\Xb$. Fisher scoring is equivalent to the Newton-Raphson method when we use the canonical link in logistic regression. Since $y_i\in(0,1)$, $-\Xb^T\Wb\Xb$ is strictly negative definite. When $y_i$ is too close to $0$ or $1$, weights are close to $0$ and $\Hb$ may have singular values close to $0$ and therefore may be computationally singular.
The Newton-Raphson algorithm \citep{Gree:1984,Clev:Devl:1988,simonoff1991improved} iterates according to
\begin{align}
\beta^{(t+1)} &= \beta^{(t)} +(\Xb^T\Wb \Xb)^{-1}\Xb^T\Wb (y-\mu)\nonumber\\
&=(\Xb^T \Wb \Xb)^{-1}\Xb^T \Wb (y-\mu+\Xb\beta^{(t)}),
\label{est_eq1}
\end{align}
which is the iteratively reweighted least squares (IRLS) algorithm \citep{Rupp:Wand:1994}. 

\textbf{Remarks}:
IRLS solves a generalized linear model in each iteration
$$
(\Xb^T \Wb \Xb)^{-1} \Xb^T \Wb z^{(t)},
$$
where $z^{(t)}=y-\mu+\Xb\beta^{(t)}$ is the residual.
When $\Wb$ is symmetric positive definite this is equivalent to the solution of the weighted least squares problem:
$$
\min_{\beta}\sum_{i=1}^n\left(\Wb^{1/2}(X_i^T \betab - y_i)\right)^2.
$$
Note that in the logistic regression case $g'(\mu) = g(\mu)(1-g(\mu))$ is never zero, so that the diagonal entries of $\Wb$ are never zero. However, other link functions might lead to degenerate cases with indefinite weighting matrices.
When the link function $g(\mu)$ is the identity function and the error follows Gaussian with mean zero and variance one ($N(0,1)$), then the IRLS algorithm becomes
\begin{align*}
    \beta^{(1)}&=0,\\
\mu&=\Xb \beta^{(1)}=0,\\
z&=\eta+\frac{y-g(\mu)}{g'(\mu)}=0+\frac{y-0}{1}=y,\\
\Wb&=\mbox{diag}\left(\frac{g'(\mu)^2}{(g'(\mu)}\right)=\mbox{diag}(1^2/1)=I_n,\\
\beta^{(2)}&=0+(\Xb^T\Xb)^{-1}\Xb^Ty,
\end{align*}
which is the ordinary least squares estimator.

\subsection{Variance heterogeneity}

When the residual errors show a trend (e.g., increasing variability as the explanatory variable increases), this can be addressed by modeling the variance as a function of the independent variable or the fitted values. Different methods such as the likelihood ratio test, score test, nonparametric test, and certain modelling approaches can be used for the diagnosis of heteroscedasticity in normal nonlinear regression models \citep{tsai1986score,simonoff1991improved,diblasi1997testing,lin2003testing}. For non-normal models, the test of departure from the nominal dispersion, including over-dispersion and under-dispersion were also discussed \citep{dean1992testing,wei1998testing}.

If variance heterogeneity is ignored, the parameter estimates may not be influenced much, but it may result in severely misleading confidence and prediction intervals \citep{carroll2017transfor}.
One way of taking into account variance heterogeneity is by explicitly
modelling it by formulating a regression model for the variance. For example, \cite{chung2007nonstationary} investigated regression models in which the conditional variance of the error is an unscaled function of an integrated time series.

The generalized least squares (GLS) method \citep{Neld:Wedd:1972} is typically used for correlated residuals with heterogeneous variances using the IRLS \eqref{est_eq1}.
Formally, we replace the assumption of the noise distribution $\epsilon\sim N(0,\sigma^2 I_n)$ with the assumption of the noise distribution as
$\epsilon\sim N(0,\Sigma)$ where $\Sigma = V^{1/2}RV^{1/2}$. Here $V$ is a diagonal matrix of potentially different (heterogeneous) variance terms and $R$ is a correlation matrix. $V$ has a constant diagonal and $R$ can be modelled by a variance function (e.g., $a_i(\phi)$ in an exponential family in \eqref{log_like}) or an autoregressive–moving-average temporal structure. Additionally, Gaussian processes \citep{rasmussen2003gaussian} with the Bayesian nonparametric inference have been widely used for heteroscedastic  nonlinear regression in which we have
$f \sim N (\mu, K)$.
The covariance matrix $K$ has elements $K_{ij}=\kappa(\xb_i,\xb_j)$, and function $\kappa$ measures similarity between two observations $\xb_i$ and $\xb_j$.
For example, the radial basis function kernel has
$
\kappa(\xb_i, \xb_j) = \exp\left\{-
\|x_i-x_j\|^2/\gamma\right\}
$ with a hyperparameter $\gamma$.

\section{Model building}
Model building aims at finding more realistic ways
to describe the stochastic behavior observed in
data. We listed a few widely used models, which are classified by their shape and applied to real data analysis in various fields \citep{Ratkowsky1983}.
\begin{itemize}
\item Polynomial models.
Linear and quadratic polynomial regression have been widely applied to solve real-world data problems for scientists and engineers. Quadratic or higher-order polynomials are curved with respect to the predictors, but they are linear in the parameters so that they can be fitted by using linear regression. However, they tend to overfit the responses with a large number of predictors.
As the availability of nonlinear regression algorithms increased, the use of polynomials has sensibly decreased. Linear or quadratic polynomials are mainly used to approximate the observed response within a small range of each predictor for understanding the relationship with the response. 
\item Concave/Convex curves models.  Concave/convex curves describe nonlinear relationships with asymptotes and without inflection points.
\begin{enumerate}
    	\item Exponential model. The exponential equation describes an increasing or decreasing trend with constant relative rate. The most common parameterization is
    	$$
    	f(x,\thetab)=\theta_1\exp(\theta_2x),
    	$$
    	which has the slope of the tangent line through $x$ is $\theta_2$. Therefore, the response increases by an amount that is proportional to $x$ positively when $\theta_1>0$ (exponential growth), and negatively when $\theta_2<0$ (exponential decay). The exponential curve is used to describe the growth of a population in unconstrained environmental conditions. The exponential function is nonlinear and can be fitted by using nls() or drm() functions in R \citep{team2018r}.
        \item Asymptotic model.
        The asymptotic regression model describes a limited growth, where $y$ approaches a horizontal asymptote as $x\to\infty$. There are several different parameterizations known as the monomolecular growth model \citep{Fekedulegn1999ParameterEO}. The most widely used parameterization is
$$
y=\theta_1-(\theta_1-\theta_2)\exp(-\theta_3x)+\epsilon, $$
where $\theta_1$ is the maximum attainable $y$, $\theta_2$ is $y$ at $x=0$, and $\theta_3$ is proportional to the relative rate of $y$ increase while $x$ increases.
        \item Negative  exponential model. This model can be given by
$$
y=\theta_1\left(1-\exp(-\theta_2x)\right),
$$
which has a similar shape to the asymptotic regression, but $y=0$ when $x=0$, and is often used to model absorbed photosynthetically active radiation \citep{kiniry1989radiation}.
        \item Power curve model.
        The power curve is also known as the Freundlich equation or allometric equation \citep{thommes2015physisorption} and the most common parameterization is
$$
y=\theta_1 x^{\theta_2}+\epsilon.
$$
This curve is equivalent to an exponential curve on the logarithm of $x$ since
$\theta_1x^{\theta_2}=\theta_1\exp\left(\log(x^{\theta_2})\right)=\theta_1\exp\left(\theta_2\log(x)\right)$, and it does not have an asymptote for $x\to\infty$.
        \item Logarithmic model. After $x$ is log-transformed, 
$$
y=\theta_1+\theta_2\log(x)
$$ is a linear model with $x > 0$. The parameter $\theta_2$ dictates the shape, as in the exponential equation, Indeed, if $\theta_2>0$, the curve is convex up and $y$ increases as $x$ increases. If $\theta_2<0$, the curve is concave up and $y$ decreases as $x$ increases.
        \item Rectangular hyperbola model.
        The Michaelis-Menten equation is a rectangular hyperbola, often parameterized as
$$
y=\frac{\theta_1x}{{\theta_2}+x}+\epsilon,
$$
which is a convex-up curve and $y$ increases as $x$ increases up to a plateau level. The parameter $\theta_1$ represents the asymptote as $x\to\infty$, while $\theta_2$ is the $x$ value giving a response equal to $\theta_1/2$. This is because $\theta_2=x_{50}$ when $\theta_1/2=\frac{\theta_1x_{50}}{{\theta_2}+x_{50}}$ where $x_{50}$ is the median.
    \end{enumerate}
    \item Sygmoidal curve model. Sygmoidal curves are S-shaped and may be increasing, decreasing, or symmetric around the inflection point. We show a common parameterization.
    \begin{enumerate}
    	\item Logistic model. The logistic curve is derived from the cumulative logistic distribution function; the curve is symmetric around the inflection point and it may be parameterized as
$$
y=\theta_1+\frac{\theta_2-\theta_1}{1+\exp(\theta_3(x-\theta_4))}+\epsilon,
$$
where $\theta_2$ is the upper asymptote, $\theta_1$ is the lower asymptote, $\theta_4$ is the $x$ value producing a response half-way between $\theta_2$ and $\theta_1$. $\theta_3$ is the slope around the inflection point, and it can be positive or negative and, consequently. Hence, $y$ may increase or decrease as $x$ increases. 
    	
        \item Gompertz model. The Gompertz curve has the following parameterization
$$
y=\theta_1+(\theta_2-\theta_1)\exp\{-\exp[\theta_3(x-\theta_4)]\}+\epsilon, 
$$
where the parameters have the same meaning as those in the logistic model, but the difference is that this curve is not symmetric around the inflection point.
    \end{enumerate}
\item Log-logistic models. The sigmoidal response curve is symmetric on the logarithm of $x$, which requires a log-logistic curve. For example, in biologic assays, the log-logistic curve is defined as follows:
$$
y=\theta_1+\frac{\theta_2-\theta_1}{1+\exp\{\theta_3[\log(x)-\log(\theta_4)]\}}+\epsilon,
$$
where the parameters have the  same meaning as the logistic model.
    \begin{enumerate}
    	\item Weibull-type 1 model.
    	The Weibull distribution is  a widely used lifetime distribution in reliability and life data analysis.
    	The type-1 Weibull curve is an alternative for the Gompertz curve like the log-logistic curve is for the logistic curve. The model function is as follows:
$$
f(x,\thetab)=\theta_1+(\theta_4-\theta_1)\{1-\exp\{-\exp[\theta_2(\log(x)-\log(\theta_3))]\}\}.
$$
The parameters have the same meaning as in the sygmoidal curves.
        \item Weibull-type 2 model.
        The type-2 Weibull curve is for the Gompertz curve what the log-logistic curve is for the logistic curve. The equation is as follows:
$$
f(x,\thetab)=\theta_1+(\theta_4-\theta_1)\exp\{-\exp[\theta_2(\log(x)-\log(\theta_3))]\},
$$
again with the same parameter interpretation as in the sygmoidal curves.
    \end{enumerate}
\end{itemize}

\section{Simulation data analysis}
We compare the nonlinear least squares (NLS) estimation with different initial values of the parameters in the nonlinear regression using a simulated dataset where the error term $\epsilon$ is not normally distributed. We show that the Bayesian approach is more appropriate to find reliable estimates as well as the confidence regions. The simulation setting is similar to the one shown in Figure 1. We use the Michaelis-Menten model~\eqref{MM-model} to simulate 100 data points with $\epsilon \stackrel{i.i.d.}{\sim} Gamma(10, 0.25)$, $ \theta_1 = 100$, and $\theta_2 = 0.05$. The covariates $\Xb$ are generated from the absolute values of $N(0, 20^2)$.\\
We implement the least squares method through the nls() function in R \citep{nls}. Two sets of starting values for the parameters $\theta_1$ and $\theta_2$ are tried, $(10, 3)$ and $(50, 0.1)$, to validate if least squares estimates can converge to the true values given bad starting values.\\
Both the estimates and $95\%$ confidence intervals for $\theta_1$ and $\theta_2$ using the least squares and Bayesian approaches are shown in Table \ref{tab:table}. 
With good initial values, both the least squares and Bayesian method give good estimates and confidence regions for the parameters. However, when the starting values are bad, the estimates are far from the true values for both methods. The 95\% confidence regions still contain the true values for Bayesian modeling, but the bands are too wide since the posterior samples tend to converge to local optimums.

\begin{center}
\begin{tabular}{ |c|c|c|c| } 
\hline
Model & Initals & $\theta_1=100$ & $\theta_2=0.05$  \\
\hline
\multirow{2}{4em}{NLS} & (10, 3) & 81.78 (73.94, 89.61) & -0.22 (-0.23, -0.21) \\ 
& (50,0.1) & 102.53 (102.38, 102.69) & 0.05 (0.05, 0.05) \\
\hline
\multirow{2}{4em}{Bayesian modeling} & (10, 3) & 84.74 (25.48, 102.67) & -0.25 (-1.18, 0.05) \\ 
& (50,0.1) & 102.53 (102.37, 102.69) & 0.05 (0.05, 0.05) \\
\hline
\end{tabular}
\captionof{table}{The estimates of $\theta_1$ and $\theta_2$ with the 95\% confidence intervals using least squares and Bayesian approach with two different settings of initial values.}
\label{tab:table}
\end{center}
\section{Further reading}
\begin{itemize}
\item Levenberg-Marquardt algorithm. The use of
the Gauss-Newton algorithm with Levenberg-Marquardt modifications \citep{levenberg1944method,marquardt:1963}
is to speed up the convergence as well
as to stabilize the computation for near-singular $\Jb^T\Jb$ in the least squares normal
equations. The Levenberg-Marquardt method compromises the Gauss-Newton method and the steepest descent
method. 

\item Multicollinearity. This occurs when some
columns in the Jacobian matrix $J$ are highly
correlated and leads to an ill-conditioned matrix of normal equations. This implies that the model may be over-parameterized, so that a simpler model or a transformation of the
predictors or parameters may be considered.

\item Nonparametric and semiparametric regression.
If the target of data analysis is to fit the response curve on the explanatory variables, a nonparametric regression could be a
better alternative than a nonlinear regression
model. However, the fitted curve may be less interpretable. Semiparametric regression which consists of both parametric and nonparametric components is an alternative to nonlinear regression modelling.

\item Kernel estimator. A one-dimensional smoothing kernel is any smooth, symmetric function $K:\R\to\R$ such that $K(x) \geq 0$ and 
$\int K(x) dx = 1,\;\int xK(x)dx = 0,\;
0<\int x^2K(x)dx <\infty.$
Given a bandwidth $h > 0$, the Nadaraya-Watson kernel regression estimator \citep{Nada:1964,Wats:1964} is defined as
$$
\hat{f}_h(x) =
\frac{\sum^n_{i=1}  K\left(\frac{\|x-x_i\|}
{h}\right)y_i}
{\sum^n_{i=1}  K\left(\frac{\|x-x_i\|}
{h}\right)}=
\sum^n_{i=1} w_i(x)y_i,
$$
where $w_i(x) = \frac{K(\|x -X_i\|/h)}{
\sum^n_{j=1}K(\|x -X_j\|/h)}
.$ Therefore, $\hat{f}_h(x)$ is a local weighted average of the $y_i$'s.
\item Initial values for parameter estimation.
For a linear model the Gauss-Newton method
finds the minimum in one iteration from any initial parameter estimates. If the nonlinear mean function is nearly linear, the convergence of the Gauss-Newton method is fast and does not depend heavily on the initial parameter estimates. However, as the magnitude of model nonlinearity is large, convergence can be slow or
even may not occur, and the resulting parameter
estimates may not be reliable. In that case, good
initial values are important.

\item Differential geometric view. The measures of
curvature described above are related to the parameterization. When a
normal error assumption is assumed, a Euclidean metric is imposed. Changing the error distribution results in a different metric and geometric structures of the distributions.  The study of probability and information using differential geometry is
called information geometry. We refer the
reader to \cite{amari2016information} for  theory and applications of information geometry.

\end{itemize}

\section*{Acknowledgement}
Hsin-Hsiung Huang is supported by the National Science Foundation grants (DMS-1924792).

\bibliography{nonlinear_reg}
\end{document}